# Asymmetric electrocaloric effects in PbSc$_{0.5}$Ta$_{0.5}$O$_3$ on field application and removal

E. Stern-Taulats[1], P. Lloveras[2], M. Barrio[2], J.-Ll. Tamarit[2],

A. Planes[3], Ll. Mañosa[3], R. W. Whatmore[4], N. D. Mathur[1] and X. Moya[1]

[1]Department of Materials Science, University of Cambridge, Cambridge, CB3 0FS, UK

[2]Departament de Física, EEBE, Universitat Politècnica de Catalunya, Av. Eduard Maristany 10-14, 08019 Barcelona, Catalonia, Spain

[3]Facultat de Física, Departament de Fisica de la Matèria Condensada, Universitat de Barcelona, Martí i Franquès 1, 08028 Barcelona, Catalonia, Spain

[4]Department of Materials, Royal School of Mines, South Kensington Campus, Imperial College London, London SW7 2AZ, UK

**Abstract**

Electrically driven thermal changes in PbSc$_{0.5}$Ta$_{0.5}$O$_3$ bulk ceramics are investigated using temperature and electric-field dependent differential scanning calorimetry and infrared thermometry. On first application and removal of electric field, we find asymmetries in the magnitude of isothermal entropy change $\Delta S$ and adiabatic temperature change $\Delta T$, due to hysteresis. On subsequent field cycling, we find further asymmetries in the magnitude of $\Delta T$ due to non-linearity in the isofield legs of entropy-temperature plots.

**Introduction**

In the last years, the search of more efficient and environmental friendly alternatives to the conventional vapour-compression cooling technology has become a rising scientific concern which has boosted the attention to the field-driven thermal phenomena around room temperature first-order phase transitions in solids [1]–[4]. In this regard, an extensive number of materials have been reported to exhibit giant

magneto-, electro-, and mechanocaloric effects, i.e. large isothermal entropy changes or adiabatic temperature changes driven by magnetic fields, electric fields, and mechanical forces, respectively. In view of the performance of a solid state refrigerator, there is particular interest in the search of materials displaying a large and persistent thermal response upon numerous field cycles [5].

Within the family of electrocaloric materials, numerous perovskite oxides have been reported to exhibit prominent thermal responses around phase transitions with large changes in the pyroelectric coefficients [6]–[9]. In this respect, materials with thin film geometry of typical thicknesses restricted to $\lesssim$ 500 nm have attracted the research interest due to their enlargement of the breakdown field and the giant thermal response that they can exhibit. However, these structures carry important drawbacks concerning their practical implementation as cooling materials, as the extremely large values of the applied electric field which are required for driving the caloric effects, or their low thermal mass, which yields low values of the heat exchange in absolute terms. For the latter concern, the design of multilayer capacitors has emerged as a feasible solution [9], [10], even though it results at the expense of adding more complexity to the structures.

The present manuscript is aimed at providing a detailed description of the electrocaloric performance in PbSc$_{0.5}$Ta$_{0.5}$O$_3$ (PST) bulk ceramics. We employ direct and quasi-direct methods based on calorimetry and infrared imaging which allow for an accurate characterization of the reversibility of the effect. In our case, the giant thermal response in the bulk ceramic of this prominent electrocaloric material is driven by modest values of the applied electric field and persists during field cycling.

PST compounds are Pb (B', B") O$_3$ perovskites in which the smaller B cations, Sc$^{3+}$ and Ta$^{5+}$, occupy the octahedral sites together with the six O$^{2-}$ anions. Within this family of perovskites, it has been shown that the alternate distribution of the B' and B" cations along their positions, i.e. their 1:1 ordering along the [111] or [1$\bar{1}$1] directions, can exist as a function of the temperature, the size difference between the B cations, and the charge difference between the B cations [11]. In the case of PST, the ordering of the B cations is close to the limit of stability [12] and the degree of order at room temperature of the same compound can be tailored by controlling the annealing time and temperature, since these ordering processes occur through nucleation of ordered domains and the corresponding growth by diffusion of the B' and B" cations [11]. The degree of order is usually parametrized by the normalized quantity $\Omega$ expressed as a function of the ratio between the intensities of the lattice and superlattice x-ray peaks, where $\Omega = 1$ stands for the completely ordered structure.

The giant electrocaloric effect in PST is exhibited around the ferroelectric (FE) – paraelectric (PE) phase transition occurring at temperatures typically between 260 and 300 K [11], [13], [14]. The high-temperature cubic perovskite structure exhibits a PE (or relaxor) behaviour which, upon cooling, transforms by a cooperative distortion of the cubic lattice to a low-symmetry polar rhombohedral structure from which spontaneous polarization emerges [15]–[18]. The general features of this phase transition are strongly related to the positional disorder, as well as other factors such as the grain size, inhomogeneities, and the presence of defects and vacancies [14]–[16], [19]–[21]. As $\Omega \rightarrow 0$ the high-temperature phase is reported to gain a relaxor character, with a more diffuse phase transition and a narrower hysteresis, a frequency-dependence of the dielectric constant in the PE state, and a weak FE hysteresis below $T_C$. The transition can be classified as second-order for $\Omega \lesssim 0.5$, where the latent heat is expected to vanish [12], [13], [22], [23]. In contrast, for PST compounds with large degrees of ordering ($\Omega > 0.8$), the first-order character of the transition is strengthened, the transition becomes sharper with a low but well-defined hysteresis [11], and increased latent heat values [12], [13]. It is worth mentioning that even though chemical disorder can be thought as a barrier for the appearance of the long-range dipole coupling responsible of the FE order, the FE phase is present in all PST compounds (for $0 \leq \Omega \leq 1$) [21].

**Experimental details**

The PST ceramic sample was prepared by milling $Sc_2O_3$ and $Ta_2O_5$ powders together and prereacting at 900 °C to form the wolframite phase $ScTaO_4$. The wolframite was then reacted with PbO at 900 °C to form a single-phase perovskite powder. The perovskite powder was then hot-pressed using $Si_3N_4$ tooling and an alumina grit packing medium at 40 MPa and 1200 °C for 6 hours.

High-resolution x-ray diffraction was performed on a 0.25 mm thick sample in transmission-mode at 0.4246 Å by means of the I15 beamline at Diamond synchrotron. Temperature control was achieved with a Cryojet 5 (Oxford Instruments) and the corresponding pattern analysis was evaluated using Topas 4.2 software.

The mass of the specimens for calorimetry and IR imaging was 240.2 and 70 mg, respectively, with a thin plate geometry of 0.45 mm thickness. The electrodes for the electric field application were prepared by covering the two principal faces of each specimen with a thin layer of silver paint. The voltage difference $\Delta V$ was applied on each specimen by joining a thin copper wire of 0.2 mm diameter on each face and connecting both wires to a 1.1 kV Keithley 2410 sourcemeter.

Calorimetric measurements under applied electric fields were performed by a custom-made DSC. Isofield and isothermal thermograms were measured at rates $dT / dt = 0.65$ K min$^{-1}$ and $dE / dt = 1.6$ kV cm$^{-1}$ min$^{-1}$. The characterization of the specific heat $c$ was done by comparison of the heat flow obtained during three calorimetric runs at 20 K min$^{-1}$ using a DSC Q2000 calorimeter from TA

Instruments with an empty pan, the PST sample, and a standard sapphire sample, respectively, as described in [24].

The IR imaging technique was carried out by means of a FLIR infrared camera, model SC7500-MB, with an image acquisition frequency of 80 Hz. The experimental setup allowed proper tracking of the sample temperature by means of a Peltier module. Electric field was cycled adiabatically, at rates of application and removal of the field of ~ 200 kV cm$^{-1}$ s$^{-1}$.

Calorimetric and thermometric measurements under applied electric fields were performed by considering the metastable states which are inherent to first-order phase transitions and following the thermal excursions protocols for a conventional caloric effect, as described elsewhere [25].

Polarization measurements were performed using a cryogenic probe fabricated in house [26]. Dielectric measurements were performed using an Agilent 4294A analyser at 0.1-100 kHz, with ramp rate ±1 K min$^{-1}$. Electrical polarization was measured on heating using a Radiant Precision Premier II with a Trek high-voltage amplifier.

**Results**

Figure 1 (a) shows the specific heat $c$ data collected on heating and cooling runs. The phase transition is first-order, as evidenced by the peaks associated with the latent heat, and their hysteresis. In temperature regions above and below the transition, $c \sim 0.3$ J g$^{-1}$. Figure 1 (b) displays the polarization as a function of temperature at different values of the applied electric field. The sudden change in polarization ($|\Delta P_0|$ = 19.4 µC cm$^{-2}$) at the same temperature interval at which the heating peak in $c$ occurs is indicative of the coupling of the FE-PE transition with the first-order phase transition. X-ray data were collected at several temperatures within the range of 240-340 K. Diffractograms confirm a structural transition at $T \sim$ 300 K from a high-temperature cubic structure with space group $Fm\bar{3}m$, to the low-temperature orthorhombic phase with space group $R3m$. As illustrated in the unit cell diagram of figure 1 (c)], the cubic $Fm\bar{3}m$ structure corresponds to the doubled basic perovskite unit cell with symmetry $Pm\bar{3}m$ due to the ordering of the $Sc^{3+}$ and $Ta^{5+}$ cations. The B-site cation order has inferred to be $\Omega \sim 0.8$, as estimated by the relative integrated peak intensities associated with the (111) and (200) lattice and superlattice planes, i.e. $\Omega^2 = \frac{\left(I_{111}/I_{200}\right)_{experimental}}{\left(I_{111}/I_{200}\right)_{calculated,\ \Omega=1}}$. Results from structural refinements are gathered in figure 1 (c), (d), and (e), which display the behaviour of the angle $\alpha$, the lattice parameter $a$, and the volume of the unit cell, respectively. In agreement with [16], [27], the change to the rhombohedral cell is detected with a sudden deviation from the $\alpha = 90°$ on cooling. As shown, the phase transition is accompanied by a modest volume change of 0.3%, which is mainly due to the particular behaviour of the lattice parameter $a$ across the transition.

The isofield thermograms measured at constant values of the applied field up to 20.0 kV cm$^{-1}$ upon heating and cooling runs are shown in figure 2 (a). The sharpness of the transition varies as the electric field is applied. In this regard, the transition becomes sharpest at 8.9 kV cm$^{-1}$, as evidenced by the pronounced peak spreading over the narrowest temperature interval of $\sim$ 10 K. Larger values of the applied electric field yield again to a more diffuse transition, which spreads over a broader temperature interval of $\sim$ 20 K. The transition temperatures shown in figure 2 (b) are determined as the peak values upon heating ($T_0^h$) and cooling ($T_0^c$). Figure 2 (b) also includes the values obtained from isothermal calorimetry discussed in more detail below. The linear fits display the sensitivity of the transition to the applied field: $T_0^h = 295.3 + 0.74\ E$ and $T_0^c = 291.8 + 0.84\ E$, which agree well with [20]. Application of the electric field energetically favours the low-temperature FE phase, thus leading to $dT_0/dE > 0$ and an electrocaloric effect of conventional nature ($\Delta S < 0$ and $\Delta T > 0$, for $\Delta E > 0$). The full baseline integration of the calorimetric peaks yields the values of the transition entropy and enthalpy changes, $\Delta S_0 = 3.0 \pm 0.1$ J K$^{-1}$ kg$^{-1}$ and $\Delta H_0 = 0.90 \pm 0.05$ J g$^{-1}$, which are in accordance with previously reported data [11], [13]. $\Delta H_0$ value is in agreement with the degree of order $\Omega$ obtained from XRD data, according to the $\Delta H_0$ vs. $\Omega$ correlation reported in ref. [12]. Within the errors, both $\Delta S_0$ and $\Delta H_0$ values remain constant over the range of applied electric fields, as illustrated in figure 2 (c) for the case of $\Delta S_0$.

Figure 2 (d) shows the entropy curves $S^c$ ($T$, $E$) (cooling) and $S^h$ ($T$, $E$) (heating) referenced [$S^c$ ($T$, $E$) = $S^h$ ($T$, $E$) = 0] at a given temperature $T_{\text{ref}} = 270$ K well below the transition. For clarity, this figure displays the particular cases at 0 and 20 kV cm$^{-1}$ for heating and cooling runs only. These curves have been computed by integrating

$$S^{c,h} = \int_{T_{\text{ref}}}^{T'} \frac{1}{T}\left(\bar{c} + \frac{dQ}{dT}\right) dT, \qquad (1)$$

where $\bar{c}$ is defined as $\bar{c} = (1-\chi)\, c^{\text{FE}} + \chi\, c^{\text{PE}}$, $\chi$ is the transformed fraction ($\chi = 1$ at the PE phase). $c^{\text{FE}}$ and $c^{\text{PE}}$ stand for the specific heat below and above the phase transitions, respectively, and are assumed to be independent of the applied electric field. The second term in the integrand of eq. (1) refers to the baseline integration of the isofield calorimetric peak at the corresponding value of the applied electric field.

Figure 2 (d) establishes the framework for the quasi-direct estimation of the electrocaloric effect and a complete characterization of its reversibility region. With this in hand, we have explored the performance of the Brayton cycle, which is predominantly proposed as a practical electrocaloric refrigeration cycle [28]–[31]. In this regard, insets in figure 2 (d) display magnified regions of the plot which incorporate illustrative diagrams of the electrocaloric Brayton cycle. Cycles in each inset depart from different starting temperatures: 302 K (top left inset) and 297 K (bottom right inset). The steps associated with the thermodynamic cycle depicted at the top left inset are described as follows: A → B, adiabatic polarization of the sample (0 → E'); B → C, isofield (E = E') heat loss transferred to the hot end of the refrigeration cycle; C → D, adiabatic depolarization (E' → 0); D → A, heat income from the cold end in the absence of electric field. In this case, the cycle is reversible and the system goes back to its initial state once the cycle has been completed (A → B → C → D → A), which contrasts with the situation depicted in the bottom right inset. The cycle departs here from a lower temperature, where the system is already in the hysteretic regime when the field is first applied in A → B. Once the electric field has been removed (C → D), the system falls on a different entropy curve and the cycle ends in state E after thermalising, instead of state A. Hence, subsequent Brayton cycles would be constrained to E → F → C → D → E.

Figure 3 (a) and (b) show illustrative measurements of the heat flow recorded under isothermal conditions upon 0 ↔ 24.4 kV cm$^{-1}$ electric field cycling. Calorimetric peaks are associated with the electric field-driven first-order transition. In each case, thermograms display the first two electric field scans which have been applied on the sample after a previous thermal excursion to temperatures well above (protocol 1, solid lines) or below (protocol 2, dashed lines) the transition either at zero field (protocol 1) or at applied field (protocol 2). In temperature intervals within the hysteretic regime, the first electric field scan ($n = 1$) drives a larger calorimetric peak which loses intensity in subsequent scans ($n > 1$). Under these conditions, the first electric field scan can drive an irreversible contribution to the caloric effect which is not driven upon subsequent scans. Figure 3 (c) displays the set of isothermal thermograms as a function of the scanning electric field at selected temperatures between 287 and 315 K. This panel shows the calorimetric signal recorded upon the first electric field scan ($n = 1$) in measurements that have followed either protocol 1 (solid lines) or 2 (dashed lines). Again, the sharpness of the transition increases for values of the applied electric between 3 and 10 kV cm$^{-1}$ and decreases for larger values. The calorimetric peak is associated with the driving values of the electric field ($E_\uparrow$ and $E_\downarrow$ for application and removal of the field, respectively). For completeness, values of the driving fields have also been included in the phase diagram presented previously in figure 2 (b). As shown, both isofield and isothermal calorimetric results are in accordance and match the same linear trends in the phase diagram. It is worth noting the reduction of the hysteresis as the applied electric field or temperature increases. This fact, together with the broadening of the calorimetric peaks at high fields shown in figure 2 (a) and figure 3 (c), suggests that the transition has a weak first-order character and that it is close to the tricritical point [11], [12], [15], [18], [27], which is expected to be at $T \sim 320$ K, $E \sim 34$ kV cm$^{-1}$ by extrapolating the linear fits of the phase diagram.

Isothermal and isofield calorimetry confer the basis for the direct and quasi-direct characterization of the electrocaloric entropy change $\Delta S$ shown in figure 3 (d). The direct estimates are computed by integration of the isothermal calorimetric peaks, i.e. $\Delta S\ (T, 0 \rightarrow E) = \int_0^E \frac{dQ}{T\ dE} dE$. The quasi-direct estimates have been computed by considering the thermal history of the sample linked to the particular measurement protocol and by subtracting the entropy curves at each field [figure 2 (d)] with respect to the reference curve at zero field (or at $E$ field). For the first scan ($n = 1$), the corresponding entropy curves are chosen in accordance with the fact that the PE $\rightarrow$ FE (PE $\rightarrow$ FE) transition can be driven by either cooling (heating) or increasing (decreasing) the applied electric field. Hence, $0 \rightarrow E$ scans are computed by use of the cooling curves, i.e. $\Delta S\ (T, 0 \rightarrow E) = S^c\ (T, E) - S^c\ (T, 0)$; and $E \rightarrow 0$ scans, as $\Delta S\ (T, E \rightarrow 0) = S^h\ (T, 0) - S^h\ (T, E)$. After the first isothermal application or release of the electric field, the system falls into a thermodynamic state with entropy values $S^c\ (T, E)$ or $S^h\ (T, 0)$, respectively. The reversible component of the electrocaloric effect can now be driven upon successive cycling if favourable energetic conditions are met, i.e. $S^c\ (T, E) < S^h\ (T, 0)$. In this reversible regime of the first-order transition, subsequent isothermal electric field scans (n > 1) lead to $|\Delta S\ (T, 0 \leftrightarrow E)| = |S^c\ (T, E) - S^h\ (T, 0)|$.

Figure 4 (a) shows illustrative IR images at different instants in which the sample has been set to adiabatic electric field cycling. Images show one of the lateral faces of the PST sample which is enclosed by the electrodes at the top and bottom sides. The sample lays horizontally on the Peltier module, which monitors the background temperature of the experiment. Temperature is read by averaging the values enclosed by the two boxes drawn in image 1 of panel (a). As shown, a small piece of black tape is stuck on the sample surface so that temperature can be calibrated in view of its high IR emissivity. The plot of the calibrated temperature as a function of time of this illustrative electric field cycling is shown in figure 4 (b). Labels indicate the instants of the snapshots presented previously in panel (a). As evidenced in panel (b), adiabatic conditions are satisfied because the thermalisation time between the sample and surroundings ($\sim 10^2$ s) is three orders of magnitude larger than the scanning rate of the electric field ($\sim 10^{-1}$ s).

The whole direct characterization of the electrocaloric temperature change is completed by cycling the sample from different starting temperatures and identifying the corresponding temperature jumps. Measurements are performed by following the thermal excursions (protocol 1 and 2) described in the previous section so that the reversible and irreversible components of the electrocaloric effect can be determined. The quasi-direct characterization of $\Delta T$ is computed by inverting the entropy curves at each field, i.e. $T^c$ (S, E) = $[S^c (T, E)]^{-1}$ and $T^h$ (S, E) = $[S^h (T, E)]^{-1}$. For the first electric field scan ($n = 1$), $\Delta T$ (S, 0 → E) = $T^c$ (S, E) − $T^c$(S, 0) and $\Delta T$ (S, E → 0) = $T^h$ (S, 0) − $T^h$(S, E). For subsequent scans ($n > 1$), $|\Delta T$ (S, 0 ↔ E)$|$ = $|S^c$ (S, E) - $S^h$ (S, 0)$|$. Instead of the entropy S at which the adiabatic process has been held, it is more convenient to report $\Delta T$ values as a function of the initial temperature at which the adiabatic temperature change is exhibited, i.e. $T^c$(0) and $T^h$(E) for $n = 1$, and $T^c$(E) and $T^h$(0) for $n > 1$. Figure 5 (c) gathers the direct (circles) and quasi-direct (lines) estimation of the adiabatic temperature change at different values of the applied field. In general, the good concordance between both methods ensures high confidence in the obtained results.

The persistence of the caloric features has been tested for $10^5$ electric field cycles (0 ↔ 13.3 kV cm$^{-1}$). Figure 6 displays the $\Delta T$ values referenced at 298 K at different time intervals during the cycling process. As shown, the electrocaloric effect remains unaffected at ~1.5 K (for 0 → 13.3 kV cm$^{-1}$) and ~1.3 K (for 13.3 kV cm$^{-1}$ → 0) during the whole cycling process.

**Discussion**

Our PST ceramic reports a large electrocaloric effect which originates in the vicinity of the room temperature first-order ferroelectric phase transition. Under these conditions, the latent heat of the transition principally contributes to the caloric effect as the transition is field-driven. According to the Clausius Clapeyron equation, the change in spontaneous polarization associated with the FE → PE transition $|\Delta P_0|$ = 19.4 µC cm$^{-2}$ [figure 1 (b)] yields a sensitivity of the transition to the electric field of $|dT_0/dE| = |\Delta P_0|/|\rho \Delta S_0|$ = 0.72 kV cm$^{-1}$ K$^{-1}$, where $\rho$ is the density calculated from the unit cell parameters presented in figure 1 (~ 9 g cm$^{-3}$). This value is in good agreement with the phase diagram obtained from calorimetric measurements [figure 2 (c)], where $|dT_0^h/dE|$ = 0.74 kV cm$^{-1}$ K$^{-1}$.

Figures 3 (c) and 5 (c) provide a complete characterization of the ECE in PST and illustrate that a reversible thermal response of giant magnitude $|\Delta S|$ (0 ↔ 24.4 kV cm$^{-1}$) = 2.8 J K$^{-1}$ kg$^{-1}$ and $|\Delta T|$ (0 ↔ 13.3 kV cm$^{-1}$) = 1.6 K is driven by particularly modest values of the applied electric field. It is shown that for $|\Delta E|$ = 24.4 kV cm$^{-1}$ the electrocaloric entropy change is close to the $\Delta S_0$ value, which can be considered as its saturation value. In this regard, the maximum adiabatic temperature change can be estimated as $\Delta H_0 / c$ ~ 3 K. Due to the conventional character of the caloric effect, the thermal effect, and the corresponding reversibility region expand towards higher temperatures as values of the applied field increase. In particular, the thermal effect becomes reversible for $T \geq$ 295 K. It has been acknowledged that phase transitions sitting close to their critical point are proper candidates to yield an optimised thermal response of enlarged magnitude and reversibility [32], [33]. In our case, we detect a reduction of the hysteresis for higher fields and temperatures due to the proximity of its phase transition tricritical point ($T \sim 320$ K, $E \sim 34$ kV cm$^{-1}$). As a consequence of that, the effect becomes more reversible if the transition is driven for larger driving fields ($E_\uparrow$ and $E_\downarrow$) or at higher temperatures. This can be appreciated in direct results shown in figure 3 (d) and figure 4 (c), which evidence large reproducibility of the first electric field scan when applied at higher temperature regions.

The maxima of the $|\Delta S|$ and $|\Delta T|$ curves obtained on the first field scan ($n = 1$) are systematically larger for negative electric field scans ($E \rightarrow 0$) rather than for positive scans ($0 \rightarrow E$). In particular, $|\Delta S|$ (20 kV cm$^{-1}$ $\rightarrow$ 0) = 2.6 J K kg$^{-1}$ > $|\Delta S|$ (0 $\rightarrow$ 20 kV cm$^{-1}$) = 2.2 J K kg$^{-1}$, $|\Delta S|$ (13.3 kV cm$^{-1}$ $\rightarrow$ 0) = 2.4 J K kg$^{-1}$ > $|\Delta S|$ (0 $\rightarrow$ 13.3 kV cm$^{-1}$) = 1.9 J K kg$^{-1}$; and $|\Delta T|$ (13.3 kV cm$^{-1}$ $\rightarrow$ 0) = 2.2 K > $|\Delta T|$ (0 $\rightarrow$ 13.3 kV cm$^{-1}$) = 1.6 K. This anomaly is attributed to the fact that the hysteresis and the broadness of the transition depend on the applied electric field and on the direction in which the transition is induced, i.e. FE $\rightarrow$ PE or PE $\rightarrow$ FE. In other words, the electrocaloric effect is induced from different metastable states which have a strong dependence on $E$, $T$ and the thermal history of the sample. Accordingly, the set of entropy curves report asymmetries that have an impact on the magnitude of the electrocaloric effect. With regards to the reversible values obtained upon successive cycling ($n > 1$), both $|\Delta S|$ and $|\Delta T|$ peaks share the same maxima magnitude irrespective of the direction in which the field is applied since $|\Delta S (T, 0 \leftrightarrow E)| = |S^c (T, E) - S^h (T, 0)|$, and $|\Delta T (S, 0 \leftrightarrow E)| = |S^c (S, E) - S^h (S, 0)|$, as expected from the first law of Thermodynamics. At this point it is worth mentioning the situation depicted in figure 6, where 13.3 kV cm$^{-1}$ $\rightarrow$ 0 adiabatic scans yield larger values of $\Delta T$ in comparison with those obtained from 0 $\rightarrow$ 13.3 kV cm$^{-1}$ scans. In this case, both processes share the same starting temperature, but the whole cycling process is not isentropic ($S_{0 \rightarrow E} \neq S_{E \rightarrow 0}$) since the sample is allowed to thermalise and exchange heat with the environment.

In view of the large reversibility of the electrocaloric response, the electrocaloric Brayton cycle has been presented in figure 2 (d). For the particular case of the cycle which departs at 302 K, the heat absorbed from the cold end is estimated as $Q_{D \rightarrow A} = \int_{S(T_D,E)}^{S(T_A,E)} T\, dS = 596$ J kg$^{-1}$. The whole cycle is performed at the expense of a net amount of work, which is calculated as the area enclosed by the cycle in the S-T diagram, i.e. $W_{A \rightarrow B \rightarrow C \rightarrow D} = \oint T\, dS = 3.5$ J kg$^{-1}$ [yellow region in figure 2 (d)]. In this regard, the coefficient of performance (COP) is defined as the ratio between the cooling effect ($Q_{D \rightarrow A}$) and the work required to drive the full cycle. For the present cycle, COP = $Q_{D \rightarrow A}$/ $W_{A \rightarrow B \rightarrow C \rightarrow D}$ ~ 170. It is worth pointing out that this value represents the upper unattainable threshold for any real electrocaloric refrigerator operating with our PST ceramic as a refrigerant being under Brayton cycling, since it assumes ideal conditions of heat transfer, work recovery, and neglects any energy loss that would be involved in the operability of the refrigerator components.

The electrocaloric performance of the PST ceramics has been tested to persist after a significant number of cycles, as evidenced by figure 5. On the one hand, our samples exhibit an outstanding dielectric behaviour, as inferred from its poor electrical conductivity ($k$ ~ 7 x 10$^{-15}$ S m) and the very low leakage currents ($I$ ~ 10 nA), which lead to negligible Joule heating. On the other hand, it is worth noting that the phase transition is accompanied by a relatively low volume change [see figure 1 (e)]. As a consequence of that, stresses between domains and grain boundaries favouring the creation and growth of cracks during cycling may be minimized. In this regard, it has been acknowledged that the proliferation of cracks, dislocations or defects favour the creation of the electric current paths which can easily yield the electrical breakdown [34], [35].

Table 1 collects significant parameters related to the ECE exhibited by our PST ceramic sample in comparison with the results from previous studies. The electrocaloric strength written as |$\Delta S$ / $\Delta E$| and |$\Delta T$ / $\Delta E$| is a convenient parameter for the comparison of the thermal response in relative terms of the

applied field. Our direct and quasi-direct results are indicative of a larger ECE in PST ceramic samples if contrasted with the study reported in [23]. The larger thermal response measured in thin films [8] is driven at the expense of giant electric fields and, thus, yielding to lower EC strength values. Overall, the EC performance of the present PST bulk ceramic can be classified among one of the largest in magnitude [1], [9] being reversibly driven for low values of the applied electric field.

**Conclusions**

Our PST ceramic exhibits a first-order phase transition with coupled structural and electrical changes (rhombohedral $R3m$ FE phase → cubic $Fm\bar{3}m$ PE phase) that set the basis for an optimized electrocaloric response.

We have analysed the giant ECE arising in the vicinity of the first-order FE → PE transition in a PST ceramic sample by means of direct and quasi-direct methods. Both methods reveal concordant results and provide a confident and complete characterization of the magnitude of the ECE and its reversibility. In PST samples with intermediate degrees of B-cation order, the transition displays a competition between the first and second-order character. Our sample has a relatively high ordering ($\Omega \sim 0.8$) and the transition is first-order, even though that it becomes more diffusive and less hysteretic for large $E$ values ($\geq 12$ kV cm$^{-1}$), which might be due to the proximity of the transition to the tricritical point. Thus, this competition first- and second-order features prompts the studied compound to display optimized features of low-hysteresis ($\sim 4$ K) and enhanced reversibility upon field cycling, which combines with the large caloric contribution associated with the latent heat ($\Delta S \sim 3$ J K$^{-1}$ kg$^{-1}$). The presented results consolidate PST among the best EC materials. The compound exhibits a giant ECE of magnitude $|\Delta S|$ (0 ↔ 24.4 kV cm$^{-1}$) = 2.8 J K$^{-1}$ kg$^{-1}$ and $|\Delta T|$ (0 ↔ 13.3 kV cm$^{-1}$) = 1.6 K that is reversibly driven by modest electric fields, which contrasts with the large values of the electric field which typically drive the thermal response in thin films.

The enlarged reversibility of the ECE in PST motivates the interest to explore the sample cyclability and its applicability to cooling purposes. On the one hand, the conditions yielding an optimised Brayton refrigeration cycle operating under 0 ↔ 24.4 kV cm$^{-1}$ electric fields are discussed and the present material is postulated to exhibit a promising electrocaloric performance under Brayton cycling. On the other hand, the ECE has been proven to remain unaffected after many electric field cycles (10$^5$) in view of the excellent conditions from both the structural and the electric points of view, as evidenced by the low volume change associated with the phase transition, the high resistivity leading to very low leakage currents and a high dielectric strength.

## Acknowledgments

E S.-T. is grateful for support from The Royal Society. X.M. is grateful for support from ERC Starting grant no. 680032 and the Royal Society.

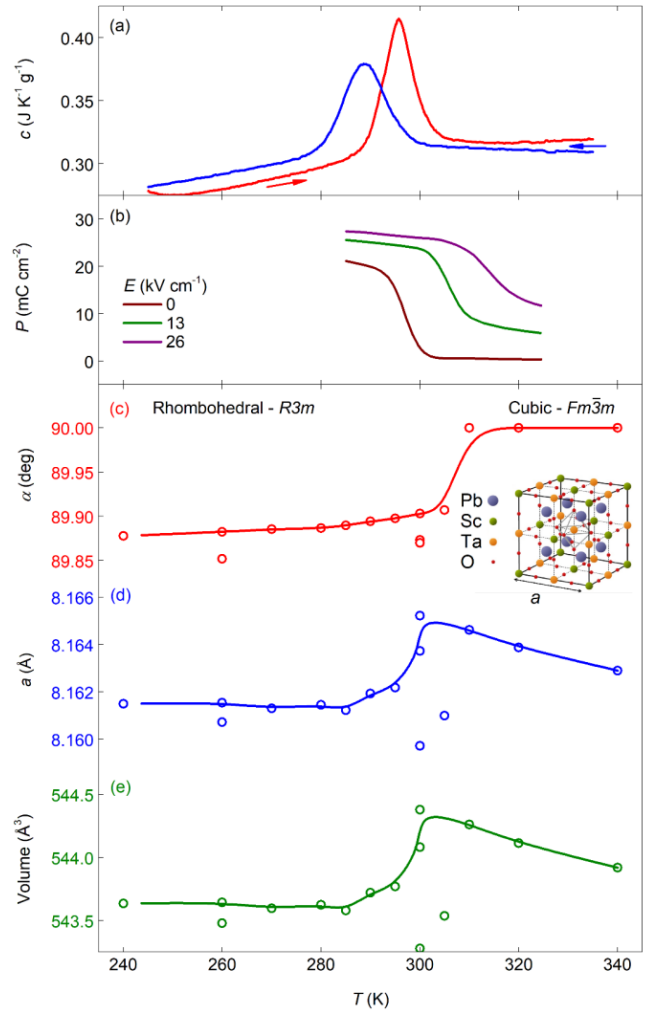

**Figure 1. Ferroelectric phase transition in PST.** Temperature-dependent (a) specific heat capacity, (b) polarisation, and (c) lattice parameters, on heating. Inset in panel (c) shows the unit cell of the fully ordered high-temperature cubic phase ($Fm\bar{3}m$).

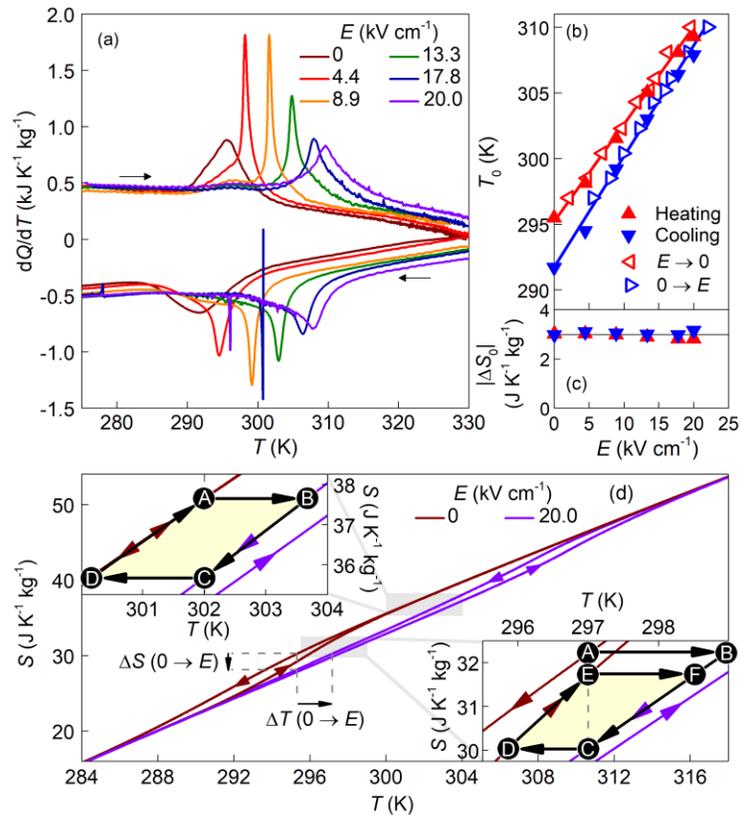

**Figure 2. Quasi-direct electrocaloric measurements of heat.** (a) Isofield d$Q$/d$T$ obtained on heating (d$Q$/d$T$ > 0) and cooling (d$Q$/d$T$ < 0) under constant applied electric field. (b) Phase diagram of the FE-PE first-order phase transition. Lines are linear fits. (c) Transition entropy change as a function of the applied electric field on heating (red triangles) and cooling (blue triangles). (d) Plot of entropy-temperature at 0 kV cm$^{-1}$ and 20 kV cm$^{-1}$. Insets in (d) show Brayton cycles with starting temperature well-above the zero-field hysteretic region (left inset, $T_s$ = 302 K), and within the zero-field thermal hysteretic region (right inset, $T_s$ = 297 K).

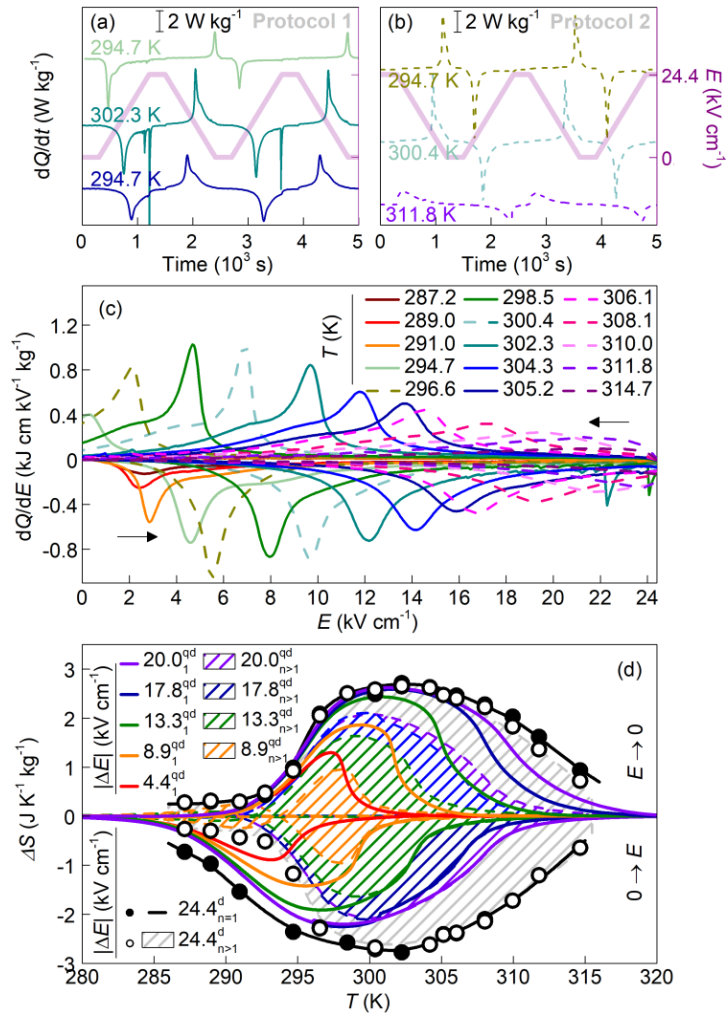

**Figure 3. Direct electrocaloric measurements of heat.** (a,b) d$Q$/d$t$ recorded under isothermal conditions upon electric field cycling. Cycles in (a) start with zero electric field (protocol 1); cycles in (b) start with 24.4 kV cm$^{-1}$ (protocol 2). (c) Corresponding d$Q$/d$E$. (d) Direct ($^\text{d}$, circles) and quasi-direct ($^\text{qd}$, lines) values of electrically driven isothermal entropy change. Solid circles correspond to the first electric field scan (n=1) and open circles to subsequent scans (n>1). Shaded coloured areas represent reversible effects.

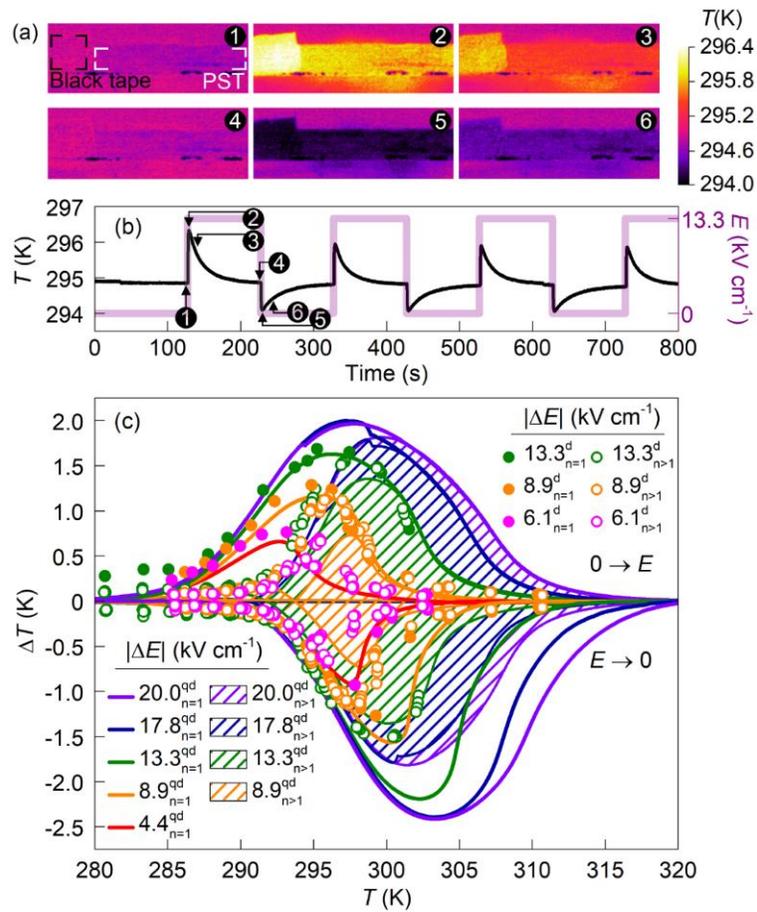

**Figure 4. Direct electrocaloric measurement of temperature change.** (a) Set of illustrative IR images [panels (1 – 6)] of the sample and the black tape portion during an electric field cycle. Legend on the right displays the colour-to-temperature conversion. (b) Calibrated temperature of the PST block upon 0 ↔ 13.3 kV cm$^{-1}$ electric field cycling, as indicated by the thick purple lines. Labels 1-6 indicate the time instants at which images shown in (a) correspond. (c) Direct ($^d$, circles) and quasi-direct ($^{qd}$, lines) estimations of the adiabatic temperature change. Solid circles correspond to the first electric field scan (n=1) and open circles to subsequent scans (n>1). Shaded coloured areas represent reversible effects.

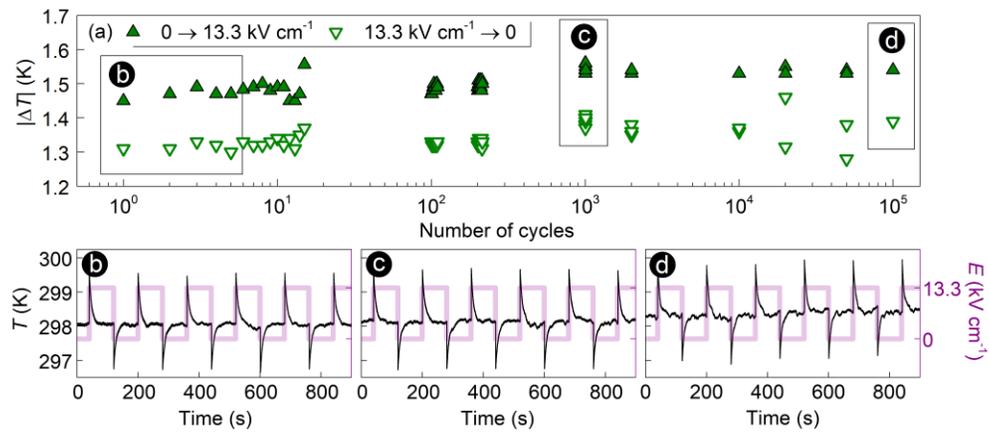

**Figure 5. Electrocaloric fatigue**. (a) Evaluation of the adiabatic temperature change upon $10^5$ electric field cycles ($0 \leftrightarrow 13.3$ kV cm$^{-1}$) $T \sim 298$ K. Panels (b, c, d) display the temperature profile at different time intervals, departing from the cycle number 1, $10^3$, and $10^5$, respectively, as indicated in (a).

| Sample | $\Omega$ | $T$ (K) | $|\Delta E|$ (kV cm$^{-1}$) | $|\Delta S|$ (J K$^{-1}$ kg$^{-1}$) | $|\Delta T|$ (K) | $|\Delta S/\Delta E|$ (mJ cm K$^{-1}$ kV$^{-1}$ kg$^{-1}$) | $|\Delta T/\Delta E|$ (mK cm kV$^{-1}$) | Ref. |
|---|---|---|---|---|---|---|---|---|
| Bulk, PC | 0.8 | 298 | 24.4 | 2.8$^d$ (2.8$^d$) | - | 110 (110) | 90 | This work |
|  | 0.8 | 298 | 20.0 | 2.6$^{qd}$ | 2.2$^{qd}$ | 130 | 110 | This work |
|  | 0.8 | 297 | 13.3 | 2.2$^{qd}$ | 1.7$^d$ (1.6$^d$) | 170 | 130 (120) | This work |
|  | 0.8 | 296 | 8.9 | 1.6$^{qd}$ | 1.3$^d$ (1.2$^d$) | 180 | 150 (130) | This work |
| Bulk, PC | 0.85 | 295 | 25 | 1.6$^c$ | 1.5$^i$ | 65 | 60 | [23] |
|  | 0.34 | 265 | 25 | 0.6$^c$ | 0.5$^i$ | 24 | 20 | [23] |
|  | 0 | 260 | 25 | 0.1$^c$ | 0.1$^i$ | 4.9 | 4 | [23] |
| Thin film (200 nm), PC | 0.32 | 337 | 774 | 6$^i$ | 6.9$^i$ | 8 | 9 | [8] |
|  | 0.32 | 337 | 201 | 1.5$^i$ | 1.5$^i$ | 2 | 2 | [8] |

**Table 1. Electrocaloric effects in PST.** Type of sample (PC: polycrystalline sample); degree of order; temperature; electric field change; EC entropy change; EC temperature change; EC strength in terms of the induced entropy and temperature per applied field. Values in parentheses refer to the corresponding reversible component upon successive field cycling. Superscripts refer to the measurement method, $^d$: direct, $^{qd}$: quasi-direct, $^i$: indirect, $^c$: calculated by approximating $\Delta S \sim \frac{c \, \Delta T}{T}$ where $c \sim 0.32$ J g$^{-1}$.